\documentclass[aps,superscriptaddress,reprint]{revtex4-1}
\usepackage{amsmath}
\usepackage{amssymb}
\usepackage{amsthm}
\usepackage{graphicx,subfigure}
\usepackage[bookmarks=false]{hyperref}
\usepackage{bm,bbm}
\hypersetup{colorlinks=true,citecolor=blue,linkcolor=red,%
urlcolor=blue,pdfstartview=FitH,bookmarksopen=true}

\usepackage[T1]{fontenc}
\usepackage{txfonts}

\DeclareMathOperator{\Tr}{Tr}

\newtheorem{theorem}{Theorem}

\newtheorem{corollary}[theorem]{Corollary}

\newcommand{\bra}[1]{\langle #1\rvert}
\newcommand{\ket}[1]{\lvert #1\rangle}

\newcommand{\abs}[1]{\lvert #1\rvert}
\newcommand{\norm}[1]{\lVert #1\rVert}

\newcommand{\id}{\mathrm{id}}
\newcommand{\tr}{\mathrm{tr}}
\newcommand{\typ}{\mathrm{typ}}
\newcommand{\atyp}{\mathrm{atyp}}

\begin{document}

\title{Flag Additivity in Quantum Resource Theories}
\author{C. L. Liu}
\affiliation{Department of Physics, Shandong University, Jinan 250100, China}
\affiliation{Institute of Physics, Beijing National
  Laboratory for Condensed Matter Physics, Chinese Academy of
  Sciences, Beijing 100190, China}
\author{Xiao-Dong Yu}
\email{Xiao-Dong.Yu@uni-siegen.de}
\affiliation{Department of Physics, Shandong University, Jinan 250100, China}
\affiliation{Naturwissenschaftlich-Technische Fakult\"at, Universit\"at Siegen,
Walter-Flex-Str. 3, D-57068 Siegen, Germany}
\author{D. M. Tong}
\email{tdm@sdu.edu.cn}
\affiliation{Department of Physics, Shandong University, Jinan 250100, China}
\date{\today}
\begin{abstract}
  Quantum resource theories offer a powerful framework for studying various
  phenomena in quantum physics. Despite considerable effort has been devoted to
  developing a unified framework of resource theories, there are few common properties
  that hold for all quantum resources. In this paper, we fill this gap
  by introducing the flag additivity based on the tensor product structure and
  the flag basis for the general quantum resources. To illustrate the
  usefulness of flag additivity, we show that flag additivity can be used to
  derive other nontrivial properties in quantum resource theories, e.g.,
  strong monotonicity, convexity, and full additivity.
\end{abstract}

\maketitle

\section{Introduction}

From quantum entanglement to quantum coherence, quantum resource theories have
been used to quantify desirable quantum effects, develop new protocols for the
resource detection, and identify processes that optimize the resource
utilization for a given application
\cite{Horodecki.etal2009,Plenio.Virmani2007,Guehne.Toth2009,Modi.etal2012b,
Streltsov.etal2017,Hu.etal2018}.

All quantum resource theories have two common fundamental ingredients: free states and
free operations
\cite{Horodecki.Oppenheim2013b,Brandao.Gour2015,Coecke.etal2016,Chitambar.Gour2018}.
For a specific quantum resource, free states are quantum states that do not
contain this kind of resource. Correspondingly, free operations can not
generate this kind of resource from free states. Based on the definition of
free states and free operations, the resource measures can be introduced. In
general, a resource measure must satisfy the nonnegativity and the
monotonicity. Other useful properties, such as the strong monotonicity, the
convexity, and the additivity, may also be introduced in different physical and
mathematical contexts
\cite{Vedral.etal1997,Vedral.Plenio1998,Popescu.Rohrlich1997,Vidal2000,
Horodecki2001,Baumgratz.etal2014,Yu.etal2016b,Aberg2006,Levi.Mintert2014,
Plenio2005,Winter.Yang2016,Vollbrecht.Werner2001}.

This common structure of quantum resource theories suggests the existence of
common properties that can be applied to the general quantum resources
\cite{Brandao.Gour2015,Gour2017}. For example, in Ref.~\cite{Brandao.Gour2015},
the authors showed that under a few physically motivated assumptions,
a resource theory is asymptotically reversible if the set of allowed operations
is maximal. However, despite considerable effort has been devoted to developing
a unified framework of resource theories, few common properties that hold for general
resource measures have been found.

In this paper, we fill this gap by introducing the notions of flag bases and
flag additivity for general quantum resources. To illustrate the usefulness of
these general properties, we show that the flag additivity implies other
nontrivial properties of resource measures in quantum resource theories, e.g.,
the strong monotonicity, the convexity, and the additivity. We find that the
flag additivity holds if and only if both the strong monotonicity and the
convexity hold, the flag additivity implies the equivalence between the
additivity and the full additivity, and for regularized resource measures the
flag additivity is equivalent to the full additivity.

The paper is organized as follows. In Sec.~II, we recall the basic framework of
quantum resource theories, and introduce the notions of flag bases and flag
additivity. In Sec.~III, we prove that the flag additivity holds if any only if
the additivity and the strong monotonicity hold. In Sec.~IV, we show that the
flag additivity implies the equivalence between the additivity and the full
additivity. In Sec.~V, we discuss the flag additivity for the regularized
resource measures, and prove that the flag additivity is equivalent to the full
additivity in this special case. We conclude in Sec.~VI.

\section{Flag additivity in quantum resource theories}

For a specific quantum resource, the set of free states, denoted by
$\mathcal{F}$, contains all the quantum states that do not contain this kind of
resource, and the set of free operations, denoted by $\mathcal{O}$, contains
the quantum operations that cannot create this kind of resource. We will use
the $2$-tuple $(\mathcal{F},\mathcal{O})$ to denote this specific quantum
resource theory. For example, in the resource theory of entanglement, the free
states $\mathcal{F}$ are all separable states, and the free operations
$\mathcal{O}$ can be chosen as all LOCC (local operations and classical
communication) \cite{Bennett.etal1996c,Vedral.etal1997}. In the
resource theory of coherence, the free states $\mathcal{F}$ are all incoherent
states, and the free operations $\mathcal{O}$ can be chosen as all incoherent
operations \cite{Baumgratz.etal2014}.

With the definitions of the free states and free operations, resource measures
can be introduced. The basic requirements for a functional $M$ being a measure
for $(\mathcal{F},\mathcal{O})$ are
\begin{enumerate}
  \item[(M1)] (Nonnegativity)
    $M(\rho)\ge 0$ for any quantum state $\rho$, and $M(\rho)=0$ if (and only
    if) $\rho\in\mathcal{F}$.
  \item[(M2)] (Monotonicity)
    $M(\rho)\ge M(\Lambda(\rho))$ for any quantum state $\rho$ and any
    $\Lambda\in\mathcal{O}$.
\end{enumerate}
To study the flag additivity of the quantum resource
$(\mathcal{F},\mathcal{O})$, we need to consider the situation of appending or
discarding an auxiliary system. For this, we recall the following tensor
product structure of $(\mathcal{F},\mathcal{O})$ (see e.g.,
Refs.~\cite{Brandao.Gour2015,Chitambar.Gour2018}).
\begin{enumerate}
  \item[(T1)]
    Appending a free state is a free operation: For any given free state
    $\delta^B\in\mathcal{H}_B$, the operation
    $\Phi_\delta(\rho^A)=\rho^A\otimes\delta^B$ is a free operation from
    $\mathcal{H}_A$ to $\mathcal{H}_{AB}$.
  \item[(T2)]
    Discarding a system is a free operation: The partial trace
    $\Tr_B(\rho^{AB})=\rho^A$ is a free operation $\mathcal{H}_{AB}$ to
    $\mathcal{H}_A$.
  \item[(T3)]
    A free operation is completely free: If $\Lambda_A$ is a free operation
    on $\mathcal{H}_A$, then $\Lambda_A\otimes \id_B$ is a free operation on
    $\mathcal{H}_{AB}$.
\end{enumerate}
For example, both the resource theory of entanglement and the resource theory
of coherence satisfy this tensor product structure \cite{Chitambar.Gour2018}.

Equipped with these notions in quantum resource theories, we can define the
flag basis and the flag additivity. A basis $\{\ket{\varphi_i}\}_{i=1}^n$ of
quantum system $\mathcal{H}$ is called a flag basis if it satisfies: (i)
$\ket{\varphi_i}$ are free states for all $i=1,2,\dots,n$; and (ii) the
projective measurement $P=\{\ket{\varphi_i}\bra{\varphi_i}\}_{i=1}^n$ is a free
operation. For example, in the resource theory of entanglement, any separable
basis is a flag basis, and in the resource theory of coherence, any incoherent
basis is a flag basis. Hereafter, we will assume the quantum resource
$(\mathcal{F},\mathcal{O})$ always satisfies the tensor product structure and
the flag basis always exists.

Consider a flagged state
$\sum_{i=1}^np_i\rho_i^A\otimes\ket{\varphi_i}\bra{\varphi_i}^B$, where
$\{\ket{\varphi_i}^B\}_{i=1}^n$ is a flag basis. As all $\ket{\varphi_i}^B$ are
free states, hence the resource is only contained in the system
$\mathcal{H}^A$, which is an ensemble $\{p_i,\rho_i\}_{i=1}^n$. A reasonable
assumption is that the resource measure $M$ satisfies the following additivity
condition,
\begin{equation}
  M\left(\sum_{i=1}^np_i\rho_i^A\otimes\ket{\varphi_i}\bra{\varphi_i}^B\right)
  =\sum_{i=1}^np_iM(\rho_i^A).
  \label{eq:flagAdditivity}
\end{equation}
If Eq.~\eqref{eq:flagAdditivity} holds for any state ensemble
$\{p_i,\rho_i\}_{i=1}^n$ and any flag basis $\{\ket{\varphi_i}^B\}_{i=1}^n$, we
call that the resource measure $M$ is flag additive.

\section{Strong monotonicity and convexity}

To show that the flag additivity is of fundamental importance in quantum
resource theories, we first prove that for any quantum resource
$(\mathcal{F},\mathcal{O})$ the flag additivity holds if and only if both the
strong monotonicity and the convexity hold.

The strong monotonicity is introduced in the situation that the experimenter is
able to post-select the multiple outcomes of a quantum measurement. Suppose
that the free operation $\Lambda(\rho^A)=\sum_{i=1}^nK_n\rho^A K_n^\dagger$ is
a general quantum measurement, then the measurement gives the outcome
$\rho_i^A$ with the probability $p_i$, where $p_i\rho_i^A=K_n\rho^A
K_n^\dagger$. If the post-selection is possible, we can further perform
different free operations to different outcomes and the final operation will
also be free. As a special case, we can add different flags $\ket{\varphi_i}^B$
to different measurement outcomes. According to (T1), the final operation
\begin{equation}
  \begin{aligned}
    \tilde{\Lambda}(\rho^A)&=\sum_{i=1}^nK_n\rho^A
    K_n^\dagger\otimes\ket{\varphi_i}\bra{\varphi_i}^B \\
    &=\sum_{i=1}^np_i\rho_i^A\otimes\ket{\varphi_i}\bra{\varphi_i}^B,
  \end{aligned}
\end{equation}
is still a free operation. Then the monotonicity condition (M2) implies that
\begin{equation}
  \begin{aligned}
    M(\rho^A)&\ge M(\tilde{\Lambda}(\rho^A))\\
    &=M\left(\sum_{i=1}^np_i\rho_i^A\otimes
    \ket{\varphi_i}\bra{\varphi_i}^B\right).
  \end{aligned}
\end{equation}
If the flag additivity in Eq.~\eqref{eq:flagAdditivity} is satisfied, then we
can get that
\begin{equation}
  M(\rho^A)\ge\sum_{i=1}^np_i M(\rho_i^A).
  \label{eq:strongMonotonicity}
\end{equation}
Equation~\eqref{eq:strongMonotonicity} is usually referred to as the strong
monotonicity of the resource measure $M$, i.e., nonincreasing of $M$ under the
selective measurement on average.

The convexity is related to the mixing of quantum states. For any ensemble of
quantum states $\{p_i,\rho^A_i\}_{i=1}^n$ in $\mathcal{H}_A$, let us consider
the auxiliary state
$\sum_{i=1}^np_i\rho_i^A\otimes\ket{\varphi_i}\bra{\varphi_i}^B$, where
$\{\ket{\varphi_i}^B\}_{i=1}^n$ is a flag basis in $\mathcal{H}_B$. Now, we
discard the quantum system $\mathcal{H}_B$, i.e.,
\begin{equation}
  \Tr_B\left(\sum_{i=1}^np_i\rho_i^A\otimes\ket{\varphi_i}\bra{\varphi_i}^B\right)
  =\sum_{i=1}^np_i\rho_i^A,
  \label{eq:convexDiscard}
\end{equation}
which is a free operation according to (T2). Thus, the monotonicity condition
(M2) implies that
\begin{equation}
  M\left(\sum_{i=1}^np_i\rho_i^A\otimes\ket{\varphi_i}\bra{\varphi_i}^B\right)
  \ge M\left(\sum_{i=1}^np_i\rho_i^A\right)
  \label{eq:convexMonotone}
\end{equation}
If the flag additivity in Eq.~\eqref{eq:flagAdditivity} is satisfied, then we
can get that
\begin{equation}
  \sum_{i=1}^np_iM(\rho_i^A)\ge M\left(\sum_{i=1}^np_i\rho_i^A\right),
  \label{eq:convexity}
\end{equation}
which is usually referred to as the convexity of the resource measure $M$,
i.e., nonincreasing of $M$ under the mixing of quantum states.

Up to now, we have shown that the flag additivity in
Eq.~\eqref{eq:flagAdditivity} is sufficient for the strong monotonicity in
Eq.~\eqref{eq:strongMonotonicity} and the convexity in
Eq.~\eqref{eq:convexity}.  In the following, we will show that it is also
necessary.

Consider that the quantum system in $\mathcal{H}_{AB}$ is the flagged state
$\sum_{i=1}^np_i\rho_i^A\otimes\ket{\varphi_i}\bra{\varphi_i}^B$, where
$\{\ket{\varphi_i}^B\}_{i=1}^n$ is a flag basis in $\mathcal{H}_B$. On the one
hand, if we perform the measurement
$\{\ket{\varphi_i}\bra{\varphi_i}^B\}_{i=1}^n$ on $\mathcal{H}_B$, which is
a free operation by the definition of flag bases, then we will get the
measurement outcomes $\rho_i^A\otimes\ket{\varphi_i}\bra{\varphi_i}^B$ with the
probability $p_i$. By using the strong monotonicity in
Eq.~\eqref{eq:strongMonotonicity}, we can obtain that
\begin{align}
  M\left(\sum_{i=1}^np_i\rho_i^A\otimes\ket{\varphi_i}\bra{\varphi_i}^B\right)
  &\ge\sum_{i=1}^np_iM(\rho_i^A\otimes\ket{\varphi_i}\bra{\varphi_i}^B)\notag\\
  &\ge\sum_{i=1}^np_iM(\rho_i^A),
  \label{eq:flagge}
\end{align}
where the second inequality follows from condition (T2). On the other hand, by
using the convexity, we can get that
\begin{align}
  M\left(\sum_{i=1}^np_i\rho_i^A\otimes\ket{\varphi_i}\bra{\varphi_i}^B\right)
  &\le\sum_{i=1}^np_iM(\rho_i^A\otimes\ket{\varphi_i}\bra{\varphi_i}^B)\notag\\
  &\le\sum_{i=1}^np_iM(\rho_i^A),
  \label{eq:flagle}
\end{align}
where the second inequality follows from condition (T1). Combining
Eqs.~\eqref{eq:flagge} and \eqref{eq:flagle}, we immediately obtain the flag
additivity in Eq.~\eqref{eq:flagAdditivity}.

The preceding results can be summarized as the following theorem.
\begin{theorem}
  For any resource measure, the strong monotonicity and the convexity is
  equivalent to the flag additivity.
  \label{thm:flagAdditivity}
\end{theorem}

We should note that, in some contexts, the strong monotonicity and the
convexity are only desired but not mandatory requirements of resource measures.
Sometimes one or both of them may not hold. In this situation, we can still
consider two conditions that are weaker than the flag additivity: the flag
supadditivity and the flag subadditivity. The flag supadditivity is defined as
\begin{equation}
  M\left(\sum_{i=1}^np_i\rho_i^A\otimes\ket{\varphi_i}\bra{\varphi_i}^B\right)
  \ge\sum_{i=1}^np_iM(\rho_i^A),
  \label{eq:flagSupadditivity}
\end{equation}
and the flag subadditivity is defined as
\begin{equation}
  M\left(\sum_{i=1}^np_i\rho_i^A\otimes\ket{\varphi_i}\bra{\varphi_i}^B\right)
  \le\sum_{i=1}^np_iM(\rho_i^A),
  \label{eq:flagSubadditivity}
\end{equation}
for any state ensemble $\{p_i,\rho_i\}_{i=1}^n$ and any flag basis
$\{\ket{\varphi_i}^B\}_{i=1}^n$. By slightly modifying the proof in
Theorem~\ref{thm:flagAdditivity}, we can get the following corollary, which can
be viewed as a refinement of Theorem~\ref{thm:flagAdditivity}.
\begin{corollary}
  For any resource measure, the strong monotonicity is equivalent to the flag
  supadditivity and the convexity is equivalent to the flag subadditivity.
  \label{cor:flagSuxadditivity}
\end{corollary}

It is worth noting that Theorem~\ref{thm:flagAdditivity} and
Corollary~\ref{cor:flagSuxadditivity} can be directly applied to entanglement
measures and coherence measures. When they are applied to coherence measures,
they give the results in Ref.~\cite{Yu.etal2016b} as a special case. Compared
with the strong monotonicity and the convexity, the flag supadditivity and flag
subadditivity are much easier to prove or disprove, because they do not involve
the Kraus operators and the structure of flagged states is much simpler than
the mixing of ensembles. We can see this simplification from an example, the
trace norm of coherence $C_{\tr}$. It is difficult to prove whether or not
$C_{\tr}$ satisfies the strong monotonicity by examining it directly
\cite{Shao.etal2015,Rana.etal2016}. However, one can easily prove that
$C_{\tr}$ violates the strong monotonicity by examining the flag additivity
\cite{Yu.etal2016b}.

\section{Additivity and full additivity}

As another application of the flag additivity, we show that it implies the
equivalence between the additivity and the full additivity. In quantum resource
theories, a resource measure $M$ is said to be additive, if it satisfies that
\begin{equation}
  M(\rho^{\otimes N})=NM(\rho),
  \label{eq:additivity}
\end{equation}
for any $\rho^{\otimes N}\in\mathcal{H}_A^{\otimes N}$, where $N$ is any
positive integer. The full additivity is a stronger condition, which is defined
as
\begin{equation}
  M(\rho\otimes\sigma)=M(\rho)+M(\sigma),
  \label{eq:fullAdditivity}
\end{equation}
for any $\rho\in\mathcal{H}_{A_1}$ and $\sigma\in\mathcal{H}_{A_2}$.

In the following, we will prove that both the additivity and the full
additivity of $M$ are equivalent to the simpler condition
\begin{equation}
  M(\rho\otimes\rho)=2M(\rho),
  \label{eq:twoAdditivity}
\end{equation}
for any $\rho\in\mathcal{H}_A$. It is obvious that
$\text{Eq.}~\eqref{eq:fullAdditivity}\Rightarrow
\text{Eq.}~\eqref{eq:additivity}\Rightarrow
\text{Eq.}~\eqref{eq:twoAdditivity}$.
In order to show that they are equivalent, we only need to prove that
$\text{Eq.}~\eqref{eq:twoAdditivity}\Rightarrow
\text{Eq.}~\eqref{eq:fullAdditivity}$.

We first consider the case that both $\rho$ and $\sigma$ are states in the
identical quantum systems $\mathcal{H}_A$, i.e.,
$\rho\otimes\sigma\in\mathcal{H}_A\otimes\mathcal{H}_A$. Consider an auxiliary
state,
\begin{equation}
  \omega=\frac{1}{2}\rho\otimes\ket{\varphi_1}\bra{\varphi_1}^B+
  \frac{1}{2}\sigma\otimes\ket{\varphi_2}\bra{\varphi_2}^B,
  \label{eq:flaggedMix}
\end{equation}
where $\{\ket{\varphi_1}^B,\ket{\varphi_2}^B\}$ is a flag basis in an auxiliary
system $\mathcal{H}_B$, then the flag additivity implies that
\begin{equation}
  M(\omega)=\frac{1}{2}M(\rho)+\frac{1}{2}M(\sigma).
  \label{eq:omegaFlagAdditivity}
\end{equation}
Further, there is
\begin{equation}
  \begin{aligned}
    \omega\otimes\omega
    &=\frac{1}{4}\rho\otimes\rho\otimes\ket{\varphi_1}\bra{\varphi_1}^B
    \otimes\ket{\varphi_1}\bra{\varphi_1}^B\\
    &+\frac{1}{4}\rho\otimes\sigma\otimes\ket{\varphi_1}\bra{\varphi_1}^B
    \otimes\ket{\varphi_2}\bra{\varphi_2}^B\\
    &+\frac{1}{4}\sigma\otimes\rho\otimes\ket{\varphi_2}\bra{\varphi_2}^B
    \otimes\ket{\varphi_1}\bra{\varphi_1}^B\\
    &+\frac{1}{4}\sigma\otimes\sigma\otimes\ket{\varphi_2}\bra{\varphi_2}^B
    \otimes\ket{\varphi_2}\bra{\varphi_2}^B,
  \end{aligned}
  \label{eq:omegaTwo}
\end{equation}
From the definition of flag bases, we can easily see that the tensor product of
flag bases is still a flag basis. Hence,
$\{\ket{\varphi_i}^B\ket{\varphi_j}^B\}_{i,j=1}^2$ is still a flag basis.
Applying the flag additivity to Eq.~\eqref{eq:omegaTwo}, we get that
\begin{equation}
  \begin{aligned}
    M(\omega\otimes\omega)
    =&\frac{1}{4}M(\rho\otimes\rho)
    +\frac{1}{4}M(\rho\otimes\sigma)\\
    &+\frac{1}{4}M(\sigma\otimes\rho)
    +\frac{1}{4}M(\sigma\otimes\sigma)\\
    =&\frac{1}{4}M(\rho\otimes\rho)
    +\frac{1}{2}M(\rho\otimes\sigma)\\
    &+\frac{1}{4}M(\sigma\otimes\sigma),
  \end{aligned}
  \label{eq:omegaTwoFlagAdditivity}
\end{equation}
where we have used the relation that
$M(\rho\otimes\sigma)=M(\sigma\otimes\rho)$, as they are the same state under
reordering $\mathcal{H}_A\otimes\mathcal{H}_A$. Combining
Eqs.~\eqref{eq:twoAdditivity}, \eqref{eq:omegaFlagAdditivity}, and
\eqref{eq:omegaTwoFlagAdditivity}, we can obtain that
\begin{equation}
  \begin{aligned}
    M(\rho\otimes\sigma)&=4M(\omega)-M(\rho)-M(\sigma)\\
    &=M(\rho)+M(\sigma),
  \end{aligned}
  \label{eq:tildeTwoCopy}
\end{equation}
which is the full additivity for $\rho,\sigma\in\mathcal{H}_A$.

We then consider the general case, $\rho\in\mathcal{H}_{A_1}$ and
$\sigma\in\mathcal{H}_{A_2}$. Suppose that $\delta_1$ and $\delta_2$ are two
free states in $\mathcal{H}_{A_1}$ and $\mathcal{H}_{A_2}$, respectively.
Consider
the states
\begin{equation}
  \tilde{\rho}=\rho\otimes\delta_2, \quad
  \tilde{\sigma}=\delta_1\otimes\sigma,
  \label{eq:auxiliaryRhoSigma}\nonumber
\end{equation}
then both $\tilde{\rho}$ and $\tilde{\sigma}$ are states in the quantum system
$\mathcal{H}_{A_1}\otimes\mathcal{H}_{A_2}$. Then, applying
Eq.~\eqref{eq:tildeTwoCopy} to $\tilde{\rho}$ and $\tilde{\sigma}$, we get that
\begin{equation}
  M(\rho\otimes\delta_2\otimes\delta_1\otimes\sigma)
  =M(\rho\otimes\delta_2)+M(\delta_1\otimes\sigma).
  \label{eq:tildeFullAdditivity}
\end{equation}
By using conditions (T1) and (T2), we can get that appending and discarding
a free state are both free operations. We immediately obtain that the resource
measure $M$ does not change for appending or discarding a free state according
to condition (M2). Hence, Eq.~\eqref{eq:tildeFullAdditivity} implies that
\begin{equation}
  M(\rho\otimes\sigma)=M(\rho)+M(\sigma),
\end{equation}
which is the full additivity for any $\rho\in\mathcal{H}_{A_1}$ and
$\sigma\in\mathcal{H}_{A_2}$. Thus, we prove that
$\text{Eq.}~\eqref{eq:twoAdditivity}\Rightarrow\text{Eq.}~\eqref{eq:fullAdditivity}$.
In summary, we get the following theorem.

\begin{theorem}
  For any resource measure, the flag additivity implies the equivalence between
  the additivity and the full additivity.
  \label{thm:additivity}
\end{theorem}

\section{Flag additivity for regularized resource measures}

In general, the converse of Theorem~\ref{thm:additivity} is not true. The
equivalence of additivity and full additivity (i.e., the full additivity
itself) does not imply the flag additivity. While it holds for an important
class of resource measures, the regularized resource measures.

Suppose $M$ is a resource measure, the regularization of $M$ is defined as
\begin{equation}
  M^\infty(\rho)=\lim_{N\to\infty}\frac{M(\rho^{\otimes N})}{N}.
  \label{eq:regularizedMeasure}
\end{equation}
The most important example of regularized resource measures is the entanglement
cost, which is the regularization of the entanglement of formation
\cite{Hayden.etal2001}.  To study the asymptotic property of $M(\rho^{\otimes
N})/N$, as $N\to\infty$, we need assume a special kind of continuity called the
asymptotic continuity \cite{Nielsen2000,Donald.etal2002,Horodecki.etal2000,
Vidal2000,Synak-Radtke.Horodecki2006}. A resource measure $M$ is said to be
asymptotically continuous if it satisfies that for all states $\rho$ and
$\sigma$ in the Hilbert space $\mathcal{H}$,
\begin{equation}
  \norm{\rho-\sigma}_\tr\to 0 \Rightarrow
  \frac{\abs{M(\rho)-M(\sigma)}}{\log_2{\dim(\mathcal{H})}}\to 0,
\end{equation}
where $\dim(\mathcal{H})$ is the dimension of the Hilbert space $\mathcal{H}$,
and $\norm{\rho-\sigma}_\tr$ is the trace distance between $\rho$ and $\sigma$
\cite{Nielsen.Chuang2000}.

For regularized resource measures, we have the following theorem.

\begin{theorem}
  Suppose $M$ is a resource measure satisfying the flag additivity and the
  asymptotic continuity. Then the regularized measure $M^\infty$ is flag
  additive if and only if it is fully additive.
  \label{thm:regularizedAdditivity}
\end{theorem}

From the definition of the regularized resource measure $M^\infty$, we can
easily see that it automatically satisfies the additivity condition defined as
Eq.~\eqref{eq:additivity}. Then, the necessity follows directly from
Theorem~\ref{thm:additivity}.

To prove the sufficiency, we only need to consider the special flagged state,
\begin{equation}
  \rho=p_1\rho_1\otimes\ket{\varphi_1}\bra{\varphi_1}
  +p_2\rho_2\otimes\ket{\varphi_2}\bra{\varphi_2},
  \label{eq:twoFlagedState}
\end{equation}
where $\rho_1$ and $\rho_2$ are any two states in a Hilbert space
$\mathcal{H}$, $p_1+p_2=1$, and $\{\ket{\varphi_1},\ket{\varphi_2}\}$ is a flag
basis. The proof can be generalized to the general flagged state.

Without loss of generality, we may assume that $p_1\le 1/2$. According to the
information theory, for any $\delta>0$ and $\varepsilon>0$, $\rho^{\otimes N}$
can be written as
\begin{equation}
  \rho^{\otimes N}=(1-\varepsilon)\rho_\typ+
  \varepsilon\rho_\atyp,
  \label{eq:typicalDecomposition}
\end{equation}
when $N$ is large enough \cite{Cover.Thomas2012,Brandao.etal2007}. In
Eq.~\eqref{eq:typicalDecomposition}, the typical part $\rho_\typ$ is defined as
\begin{equation}
  \rho_\typ=\frac{1}{T}\sum_{k=\lfloor Np_1(1-\delta)\rfloor}^{\lceil
  Np_1(1+\delta)\rceil}
  \mathcal{S}\left((\rho_1\otimes\ket{\varphi_1}\bra{\varphi_1})^{\otimes
  k}\otimes(\rho_2\otimes\ket{\varphi_2}\bra{\varphi_2})^{\otimes(N-k)}\right),
  \label{eq:typicalPart}
\end{equation}
where $T$ is a normalization factor and $\mathcal{S}$ is the symmetrized tensor
product. For example,
$\mathcal{S}(\tilde{\rho}_1\otimes\tilde{\rho}_1\otimes\tilde{\rho}_2)
=\tilde{\rho}_1\otimes\tilde{\rho}_1\otimes\tilde{\rho}_2
+\tilde{\rho}_1\otimes\tilde{\rho}_2\otimes\tilde{\rho}_1
+\tilde{\rho}_2\otimes\tilde{\rho}_1\otimes\tilde{\rho}_1$,
where $\tilde{\rho}_i=\rho_i\otimes\ket{\varphi_i}\bra{\varphi_i}$. As
$k\ge\lfloor Np_1(1-\delta)\rfloor$, we can get that $N-k\le N-\lfloor
Np_1(1-\delta)\rfloor=\lceil N-Np_1(1-\delta)\rceil\le\lceil
Np_2(1+\delta)\rceil$, where we have used the condition that $p_1\le 1/2$.
Similarly, we can obtain that $N-k\ge N-\lceil Np_1(1+\delta)\rceil=\lfloor
N-Np_1(1+\delta)\rfloor\ge \lfloor Np_2(1-\delta)\rfloor$. Thus, there is
$\lfloor Np_2(1-\delta)\rfloor\le N-k\le\lceil Np_2(1+\delta)\rceil$. Then,
conditions (T2) and (M2) imply that
\begin{equation}
  \begin{aligned}
    M(\rho_1^{\otimes\lfloor Np_1(1-\delta)\rfloor}&\otimes\rho_2^{\otimes
    \lfloor Np_2(1-\delta)\rfloor})\le M(\rho_1^{\otimes
    k}\otimes\rho_2^{\otimes (N-k)})\\
    &\le M(\rho_1^{\otimes\lceil Np_1(1+\delta)\rceil}\otimes\rho_2^{\otimes
    \lceil Np_2(1+\delta)\rceil}).
  \end{aligned}
  \label{eq:subsupAdditivity}
\end{equation}

From the definition of flag bases, we can easily see that the tensor product of
flag bases is still a flag basis. Thus, by combining the flag additivity of $M$
and Eq.~\eqref{eq:subsupAdditivity}, we can get that
\begin{equation}
  \begin{aligned}
    M(\rho_1^{\otimes\lfloor Np_1(1-\delta)\rfloor}&\otimes\rho_2^{\otimes
    \lfloor Np_2(1-\delta)\rfloor})\le M(\rho_\typ)\\
    &\le M(\rho_1^{\otimes\lceil Np_1(1+\delta)\rceil}\otimes\rho_2^{\otimes
    \lceil Np_2(1+\delta)\rceil}).
  \end{aligned}
  \label{eq:asymptoticAdditivity}
\end{equation}
Let $m_1$, $m_2$, and $m$ be positive integers such that
\begin{equation}
  p_1(1-\delta)=\frac{m_1}{m}+\varepsilon_1,\quad
  p_2(1-\delta)=\frac{m_2}{m}+\varepsilon_2,
  \label{eq:approxp}
\end{equation}
where the positive parameters $\varepsilon_1$ and $\varepsilon_2$ converge to
zero when we choose suitable and large enough integers, $m_1$, $m_2$, and $m$.
Choose $N$ to be $N'm$, where $N'$ are also large enough integers. Then, the
first inequality in Eq.~\eqref{eq:asymptoticAdditivity} implies that
\begin{equation}
  M(\rho_\typ)\ge M(\rho_1^{\otimes N'm_1}\otimes\rho_2^{\otimes N'm_2}).
  \label{eq:approxSupadditivity}
\end{equation}
When $\varepsilon\to 0$ ($N\to\infty$), $\norm{\rho^{\otimes N}-\rho_\typ}\to
0$. Hence, the asymptotic continuity implies that
\begin{equation}
  \lim_{N\to\infty}\frac{1}{N}M(\rho^{\otimes N})
  =\lim_{N'\to\infty}\frac{1}{N'm}M(\rho^{\otimes N'm})
  =\lim_{N'\to\infty}\frac{1}{N'm}M(\rho_\typ).
  \label{eq:asymtoticContinuity}
\end{equation}
Combining Eqs.~\eqref{eq:approxSupadditivity} and \eqref{eq:asymtoticContinuity},
we have
\begin{equation}
  M^\infty(\rho)\ge\frac{1}{m}M^\infty(\rho_1^{m_1}\otimes\rho_2^{m_2}).
\end{equation}
Then if $M^\infty$ is fully additive, we have
\begin{equation}
  M^\infty(\rho)\ge\frac{m_1}{m}M^\infty(\rho_1)
  +\frac{m_2}{m}M^\infty(\rho_2).
  \label{eq:approxSupFlagAdditivity}
\end{equation}
Let $\varepsilon_1,\varepsilon_2\to 0$ ($m_1,m_2,m\to\infty$), we get the flag
supadditivity of $M^\infty$,
\begin{equation}
  M^\infty(\rho)\ge p_1M^\infty(\rho_1)+p_2M^\infty(\rho_2).
  \label{eq:asymptoticFlagSupAdditivity}
\end{equation}
Similarly, we can get the flag subadditivity of $M^\infty$ from the second
inequality in Eq.~\eqref{eq:asymptoticAdditivity},
\begin{equation}
  M^\infty(\rho)\le p_1M^\infty(\rho_1)+p_2M^\infty(\rho_2),
  \label{eq:asymptoticFlagSubAdditivity}
\end{equation}
when $M^\infty$ is fully additive. Combining
Eqs.~\eqref{eq:asymptoticFlagSupAdditivity} and
\eqref{eq:asymptoticFlagSubAdditivity}, we get the flag additivity,
\begin{equation}
  M^\infty(\rho)=p_1M^\infty(\rho_1)+p_2M^\infty(\rho_2).
  \label{eq:asymptoticFlagAdditivity}
\end{equation}
Thus, this completes the proof of the Theorem~\ref{thm:regularizedAdditivity}.

When Theorem~\ref{thm:regularizedAdditivity} is applied to the entanglement
measures, we immediately get the equivalence between flag additivity and full
additivity for regularized entanglement measures. This is a generalization of
the result in Ref.~\cite{Brandao.etal2007}, where an additional condition (the
subadditivity) is assumed.

\section{Conclusions}

In this paper, we have introduced the notion of the flag basis and defined the flag
additivity for general quantum resources. To illustrate the usefulness of the
flag additivity, we have shown that it can be used to derive other nontrivial
properties in quantum resource theories. As examples, we have proved that the flag
additivity holds if and only if both the strong monotonicity and the convexity
hold at the same time, the flag additivity implies the equivalence between the
additivity and the full additivity, and for regularized resource measures the
flag additivity is equivalent to the full additivity. We think the technique of
flag additivity together with its refinements, flag subadditivity and flag
supadditivity, will become a fundamental tool for studying the resource
measures.

\begin{acknowledgments}
 C.L.L. acknowledges support from the National Natural Science Foundation of
 China through Grant No. 11575101 and the National Key Research and Development
  Program of China (2016YFA0300603). X.D.Y. acknowledges support from
  the DFG and the ERC (Consolidator Grant 683107/TempoQ).
  D.M.T. acknowledges support from the National
  Natural Science Foundation of China through Grant No. 11775129.
\end{acknowledgments}


%

\end{document}